\documentstyle[12pt]{article}
\begin{document}

%       The following are some macros:

        \newcommand{\be}{\begin{equation}}
        \newcommand{\ee}{\end{equation}}
        \newcommand{\ba}{\begin{eqnarray}}
        \newcommand{\ea}{\end{eqnarray}}
        \newcommand{\ban}{\begin{eqnarray*}}
        \newcommand{\ean}{\end{eqnarray*}}
        \newcommand{\barr}{\begin{array}}
        \newcommand{\earr}{\end{array}}

        \renewcommand{\H}{{\cal H}}
        \newcommand{\K}{{\cal K}}

        \newcommand{\et}{\hspace{-0.08in}{\bf .}\hspace{0.1in}}
        \newcommand{\tensor}{\otimes}
        \newcommand{\maps}{\colon}
        \newcommand{\vol}{{\rm vol}}
        \newcommand{\iso}{\cong}
        \newcommand{\tr}{{\rm tr}}
        \newcommand{\sign}{{\rm sign}}
        \newcommand{\M}{{\it M }}
        \newcommand{\id}{{\rm id}}
        \newcommand{\Diff}{{\rm Diff}}
        \newcommand{\End}{{\rm End}}
        \newcommand{\V}{{\rm Vect}}
        \newcommand{\Vect}{{\rm Vect}}
        \newcommand{\Aut}{{\rm Aut}}
        \newcommand{\Ad}{{\rm Ad}}
        \newcommand{\Maps}{{\rm Maps}}
        \newcommand{\hf}{{1\over 2}}
        \renewcommand{\j}{\jmath}
%\renewcommand{\i}{\imath}

%      	 \textwidth 6in
%        \textheight 8.5in
%        \evensidemargin .25in
%        \oddsidemargin .25in
%        \topmargin .25in
%        \headsep 0in
%        \headheight 0in
%        \footskip .5in

%If you want single spaced copy, delete the next two lines.
       \parskip 1.75\parskip plus 3pt minus 1pt
        \pagestyle{plain}
        \pagenumbering{arabic}

\def\nn{\nonumber}
\def\bs{\bigskip}
\def\no{\noindent}
\def\hb{\hfill\break}
\def\qq{\qquad}
\def\bl{\bigl}
\def\br{\bigr}
\def\ra{\rightarrow}
\def\Ra{\Rightarrow}
\def\lra{\leftrightarrow}

\def\k{\kappa}
\def\r{\rho}
\def\a{\alpha}
\def\b{\beta}
\def\B{\Beta}
\def\g{\gamma}
\def\G{\Gamma}
\def\d{\delta}
\def\D{\Delta}
\def\e{\epsilon}
\def\c{\chi}
\def\th{\theta}
\def\Th{\Theta}
\def\m{\mu}
\def\n{\nu}
\def\na{\nabla}
\def\om{\omega}
\def\Om{\Omega}
\def\l{\lambda}
\def\L{\Lambda}
\def\s{\sigma}
\def\S{\Sigma}
\def\p{\phi}
\def\P{\Phi}
\def\t{\tau}
\def\varp{\varphi}
\def\pa{\partial}
\def\pap{\partial_+}
\def\pam{\partial_-}
\def\papm{\partial_{\pm}}
\def\pamp{\partial_{\mp}}
\def\pat{\partial_{\tau}}
\def\J{{\cal J}}

\def\frac#1#2{{#1\over#2}}
\def\pd#1#2{{\frac{\partial#1}{\partial#2}}}

        \newcommand{\Z}{{\bf Z}}
        \newcommand{\Q}{{\bf Q}}
        \renewcommand{\div}{\nabla\cdot}
        \newcommand{\curl}{\nabla\times}
        \newcommand{\grad}{\nabla}
        \newtheorem{ex}{Example}
        \newcommand{\bex}{\begin{ex}\et}
        \newcommand{\eex}{\end{ex}}
        \def\dia{\diamond}
        \def\we{\wedge}
        \newcommand{\ed}{\end{document}}

        \def\sqr#1#2{{\vcenter{\vbox{\hrule height.#2pt
                \hbox{\vrule width.#2pt height#1pt \kern#1pt
                   \vrule width.#2pt}
                \hrule height.#2pt}}}}

        \def\square{\mathchoice\sqr34\sqr34\sqr{2.1}3\sqr{1.5}3}

\font\tenmsb=msbm10 scaled\magstep{1.5}
\newfam\msbfam
\textfont\msbfam=\tenmsb
\def\Bbb#1{{\fam\msbfam\relax#1}}

\newcommand{\A}{{\Bbb {A}}}
\newcommand{\Op}{{\Bbb {O}}}
\newcommand{\lap}{{{}_{{}_{{}_{\sim}}} \!\!\!\!N}}
\newcommand{\R}{{\rm {I\!R}}}
\def\E{{\widetilde{E}}}
\def\doble{{{\lim_{{d\to 0}\atop{\e\to 0}}}}}
\def\N{{\;\lap (x)\;}}
\def\C{{\cal C}}
\def\x{{\vec x}}
\def\y{{\vec y}}

\newcounter{figg}
\newcommand{\figura}[3]{{{\begin{center}
\mbox{\epsfig{file={#1},height={#2cm}}}\end{center}
\stepcounter{figg}}{\centerline{ {\small{Fig.~\thefigg. #3}}}}\vskip 1em}}

%%%%%%%%%%%%%%%%%%%%%%%%%%%%%%%%%%%%%%%%%%%%%%%%%%%%%%%%%%%%%%%%%%%%%%
%
%	Begin of latex file
%
%%%%%%%%%%%%%%%%%%%%%%%%%%%%%%%%%%%%%%%%%%%%%%%%%%%%%%%%%%%%%%%%%%%%%

	\begin{center}
	{\Large Anomalous Commutators \\
	for  \\
	Energy-Momentum Tensors \\ }
        \vspace{1cm}
	Javier P.\ Muniain$\,{}^{\dagger}$ and Dardo D.\
P\'{\i}riz$\,{}^{\ddagger}$ \\
	\vspace{0.8cm}
	{\small{\it Department of Physics  \\
        University of California, Riverside \\
        California 92521-0413, U.\ S.\ A.  \\ }}
        \vspace{0.9cm}
	{\small May 2, 1995 \\}
	\vspace{0.9cm}
	\end{center}

\begin{abstract}
Anomalous contributions to the energy-momentum commutators
are calculated for even dimensions, by using a non-perturbative
approach that combines operator
product expansion and Bjorken-Johnson-Low limit techniques.
We first study the two dimensional case and give the
covariant expression for the commutators. The expression
in terms of light-cone coordinates is then
calculated and found to be in perfect agreement with the results
in the literature. The particular scenario of the light-cone frame
is revisited using a reformulation of the BJL limit in such a frame.
The arguments used for $n=2$ are then generalized to the case of
any even dimensional Minkowskian spacetime and it is shown that there are no
anomalous contributions to the commutators for $n\not=2$. These
results are found to be valid for both fermionic and bosonic
free fields. A generalization of the BJL-limit is later used to obtain
double commutators of energy-momentum tensors and to study the Jacobi
identity. The two dimensional case is  studied and we find no existance
of 3-cocycles in both the Abelian and non-Abelian case.
\end{abstract}

{\footnotesize ${}^{\dagger}$muniain@ucrph0.ucr.edu \qq
${}^{\ddagger}$dardo@ucrph0.ucr.edu  \\ }
\vfill\eject

\begin{center}
\subsection*{I. Introduction}
\end{center}

The evaluation of commutators and their algebra has been a subject
of interest in field theory for many years. Several approaches
have been proposed to carry out such a task, from the canonical to the
perturbative one. The use of canonical commutators to
evaluate current algebra relations has produced many results
which can be explicitly measured.
Still, in many instances, the canonical
evaluation of current commutators is ill-defined as it was shown by
Schwinger in \cite{Schw}. In that case the evaluation of the equal
time commutator $[J^a_0(x), J^b_i(y)]$ leads to the canonical result
plus a term proportional to the gradient of the delta function $S^{ab}_{ij}(y)
\pa^j \d(\vec x -\vec y)$, and possibly other higher derivatives of
delta functions, which may be present. These gradient terms are called
Schwinger terms. As canonical evaluation of equal time commutators present
ambiguities, it is necessary to have an alternative way to define and
calculate these commutators. Bjorken, Johnson and Low \cite{BJL}
proposed a definition that preserves all desirable features of the
theory. The commutator is well defined and the results obtained coincide
with the canonical ones whenever the latter are also well defined.
As we will see in the next section, this definition relies on
studying the limit of large energy transfered and therefore an
operator product expansion (OPE) \cite{Wilson} is appropiate, allowing
us to significatively reduce our calculations. The BJL-limit approach
has already been used in the literature, but always within perturbation
theory framework. These calculations turn out to be usually very lengthy
and tedious. The need to consider many loop-diagrams, with ever increasing
number of vertices, makes the study of anomalous commutators
very hard when one increases the dimensionality
of spacetime $n$. For instance, the study of anomalous commutators
between vector and axial-vector currents in $n=4$ dimensions
involves the perturbative calculation of triangle, box and even pentagon
diagrams \cite{Jo}. For the case of energy-momentum commutators,
the ones we are concerned with in this paper, the perturbative
scheme is the same. The axial or vector-axial
vertices found in the context of chiral anomalies (e.g. triangle
diagram) are replaced by energy-momentum tensors, and one can start
evaluating loop diagrams. However, instead of
doing the perturbative calculation, an OPE will show to be more efficient
and still valid for any value of the spacetime dimension.

\bs
The paper is structured in the following manner. In section II we describe
the method in detail. Section III presents the procedure to calculate in
covariant form the anomalous energy-momentum contributions to the
2-dimensional commutators. The
study is later on generalized in section IV to any even dimension. In
section V we obtain the expression of the commutator using BJL
and OPE techniques in the light-cone frame. Section VI is
dedicated to the study of the Jacobi identity using the same techniques
developed in  previous sections.

\begin{center}
\subsection*{II. Description of the method}
\end{center}

In this section we will make use of a method previously used in
\cite{mine1} within the context of non-Abelian current algebra, to obtain
the commutators of energy-momentum tensors. This method allows us
to do the calculations without going through the lengthy steps of
solving loop diagrams in perturbation theory. The method is consistent
and has the advantage that can be used for both fermionic and bosonic
models. Furthermore, results can easily be obtained for any value of the
dimensionality of spacetime. This technique, which does not rely on
perturbation theory, is based on the Bjorken-Johnson-Low (BJL)
definition of equal time commutators and on the operator product
expansion (OPE).

The BJL definition is totally general and
arises from a time ordered product of two operators, say $A$ and $B$, and
its representation in momentum space through a Fourier transform.
The BJL prescription tells us that
the equal time commutator of two operators is obtained from the high
energy behavior of Green's functions, as
\ba    && \lim_{p^0 \to \infty} p^0 \int d^n x \;e^{i p x} \langle \a |
T A(x/2) B(- x/2) | \b \rangle =    \nn		\\
&& \qq \qq \qq i \int d^{n-1} x\; e^{-i {\vec p}\cdot \x} \langle\a |
[A(0,\x/2), B(0,-\x/2)] | \b  \rangle. \label{1}
\ea
Here $p^0$ stands for the energy component of the four momentum.
In the BJL definition (\ref{1}) above we have used the time ordered
product $T$, which is not a Lorentz covariant object.
In field theory one calculates (e.g. Feynman diagrams in
perturbation theory) only the covariant object, denoted by $T^*$. The
difference
between $T$ and $T^*$ is local in position space, and corresponds to a
covariant term involving delta functions of $x_0$ and its derivatives
\cite{Jackiwbook}. This extra term will take the form of a
polynomial in $p^0$ when we go to momentum space. Therefore
in eq.(\ref{1}) we must drop all polynomials in $p^0$, since they will not
contribute to the covariant commutator. The point here is to make a Laurent
expansion of the time ordered product and identify the residue of
the $1/{p^0}$ term as the Fourier transform of the commutator.

Since we are interested in the large momentum transfered behavior, it
is appropiate
to express the singularities of the product of operators as a sum of
non-singular local operators \cite{Wilson},
\be       \int d^n x \;e^{i p x} \langle \a |T^* A(x/2) B(-x/2)| \b
\rangle = \sum_i c_i(p)\; \langle \a |{\cal O}_i (0)|\b \rangle,
\label{2}
\ee
where these local operators ${\cal O}_i$ are evaluated at $x=0$. Taking
the BJL limit in eq.(\ref{2}) we find
\be      \int d^{n-1} x\; e^{-i{\vec p}\cdot \x} \langle\a | [A(0, \x/2),
 B(0,-\x/2)] | \b \rangle = \sum_i \lim_{p^0 \to\infty} [-i p^0 c_i(p)]
\langle \a | {\cal O}_i (0)| \b \rangle,    \label{3}	\ee
where all terms in the coefficients $c_i(p)$ that grow as a power
of $p^0$ must be dropped.

\bs
\begin{center}
\subsection*{III. Commutators in two dimensions}
\end{center}

The simultaneous use of OPE and BJL techniques
to calculate commutators in a non-perturbative manner has been proven a
useful way to evaluate different kind of such operators. This method
avoids the tedious work usually associated with perturbative calculations.

In \cite{mine1} it was shown that these techniques combined together
in the context of current commutators
reproduce the results previously found in the literature \cite{Jo}.
It is our purpose now to examine the possible anomalous (non-canonical)
contributions to the commutators of energy-momentum tensors in
two dimensions. At this time we will only carry out our discussion in
a Minkowskian spacetime with metric $\eta_{\a\b}$ and signature $(+,-)$.

The two-point function for the energy-momentum
tensor is given by
\be   {\cal O}_{\m\n\r\s}(p) = \int d^2 x\;e^{ipx} \langle \Om |
T^* (\th_{\m\n} (x/2) \th_{\r\s}(-x/2)) |\Om \rangle,
\label{4}   \ee
where $\Om$ denotes the vacuum state (the method holds for
any two states $|\a \rangle$ and $|\b \rangle$).

In an n-dimensional spacetime, simple dimensional analysis tells us
that the canonical dimension of $\cal O$ is $({\rm mass})^n$. This is
straightforward if one recalls that $\lbrack \th_{\a\b}\rbrack =
\lbrack \eta_{\a\b} {\cal L} \rbrack = n$. In our present case
$\lbrack {\cal O} \rbrack = 2$, restricting the possible terms
in the OPE to those of
dimension equal or greater than two. To choose the terms to appear in
the OPE, we pick those with the same dimensionality and symmetries (Lorentz
covariance, gauge invariance, parity, etc.) as $\cal O$.
Out of all possible terms, we shall write only those that contribute
when taking the BJL limit.

To illustrate our result below, we consider the case of the
energy-momentum tensor for free fermions moving in a flat background,
used by Alvarez-Gaum\'e and Witten in \cite{witten}
\be
\th_{\a\b} = {i\over 4} \bar\psi (\g_\a {\stackrel{\lra}{\pa}}_\b + \g_\b
{\stackrel{\lra}{\pa}}_\a)\psi,  \label{4a}
\ee
which is symmetric, $\th_{\a\b} = \th_{\b\a}$, and conserved, $\pa^\a
\th_{\a\b}= 0$.
Another possibility was studied by Guadagnini in \cite{guadagnini} for the
case of a free bosonic model. In that case, the traceless and symmetric
energy-momentum tensor was given by
\be
\th_{\a\b} = \pa_\a \phi^i \pa_\b \phi^i - \eta_{\a\b} \pa_\l \phi^i
\pa_\l \phi^i.  \label{4b}
\ee

\bs
Since we are working with symmetric tensors in (\ref{4a}) and (\ref{4b}),
we want our terms in the OPE of (\ref{4}) to be symmetric when
$\m \lra \n$ and
$\r \lra \s$, separately. From the symmetries of two-point
functions we also need the following symmetry for $ \cal O$:
$(\m,\n)\lra (\r,\s)$
and $p\lra -p$. The OPE for contributing terms will be of the form
\ba    {\cal O}_{\m\n\r\s} &=&  {1\over {p^2}}[a_1 p_\m p_\n p_\r p_\s {\bf 1}
+ a_2 (p_\m p_\n \th_{\r\s} + p_\r p_\s \th_{\m\n}) + {{a_3}\over{p^2}}
p_\m p_\n p_\r p_\s \;\Th],   \label{5}
\ea
where $\Th=\th_\l^\l$ stands for the trace of the energy momentum tensor.
There will be cases where $\th_{\m\n}$ is traceless and hence the last term in
the operator expansion will not appear. When considering interaction
with the gravitational field there appears the so called
trace anomaly, where $\Th$ is proportional to the Ricci scalar $R$.

Applying the BJL limit to eq.(\ref{5})
\[	\lim_{p^0\to\infty} p^0 {\cal O}_{\m\n\r\s}= i \int d x \,
e^{-i {\vec p} \cdot \x}\; [ \th_{\m\n}(0, \x/2), \th_{\r\s}(0, -\x/2) ],    \]
we find
\ba
[ \th_{\m\n}(0, \x/2), \th_{\r\s}(0, -\x/2) ] &=& a_1\;\eta_{\m (0}\;
\eta_{\n i}\;\eta_{\r j}\;\eta_{\s k)}\; \pa^i \pa^j \pa^k \d(\vec x) \nn   \\
&-& a_2\; [\eta_{\m (0}\;\eta_{\n i)} \th_{\r\s} + \eta_{\r (0}\;
\eta_{\s i)} \th_{\m\n}]\;\pa^i\d(\vec x)  \nn   \\
&-& a_3\; \eta_{\m (0}\;\eta_{\n 0}\;\eta_{\r 0}\;\eta_{\s i)}\;\Th\;\pa^i
\d(\vec x),    \label{6}
\ea
where we have used the parentheses to denote symmetric permutation of
second indices; i.e. $\eta_{\m (0}\;\eta_{\n i)} = \eta_{\m 0}\;
\eta_{\n i} + \eta_{\m i}\;\eta_{\n 0}$.
The above expression for the commutator is Lorentz covariant and the
different components can be obtained by giving values zero and one to
$\m$, $\n$, $\r$ and $\s$. The equal time
commutators are
\[ [\th_{00}, \th_{00}] = [ \th_{00}, \th_{11}] = [\th_{11},\th_{11}]= 0,  \]
and
\ba
\lbrack \th_{00} (0, \x/2), \th_{01} (0, -\x/2) \rbrack &=& [a_2 \th_{00} +
a_3 \Th ] \d'(\vec x),  \nn   \\
\lbrack \th_{01} (0, \x/2), \th_{01} (0, -\x/2) \rbrack &=&  2 a_2 \th_{01}
\d'(\vec x),  \nn   \\
\lbrack \th_{01} (0, \x/2), \th_{11} (0, -\x/2) \rbrack &=& a_2 \th_{11}
\d'(\vec x) - a_1 \d'''(\vec x).       \label{7}
\ea
At this time we are not interested in the concrete values of the
multiplicative constants $c_i$. These constants multiply the matrix
elements $\langle \a | {\cal O}_i | \b \rangle$, which in most
cases can only be evaluated doing detailed calculations.
In expression (\ref{7}) we see that the only anomalous contribution
takes place when we have one time-like index and the other three
space-like indices (e.g. $[\th_{01},\th_{11}]$). The anomaly
appears as a third derivative of the delta function. This result is similar
to the Virasoro anomaly found for two dimensional world sheets in the context
of superstring theory \cite{super}. Also similar results are given in
\cite{witten} using light-cone formalism, which we will review in section V.

\begin{center}
\subsection*{IV. Commutators in n dimensions}
\end{center}

The previous results can be generalized to the case of an n-dimensional
spacetime, with $n$ being
even\footnote{In the odd dimensional case, some study has been carried out
for n=3 in the context of current algebra \cite{mine2}.}.
The following study will show that there are no anomalous contributions
to the commutator of energy-momentum tensors for $n \ge 4$.
For a term to contribute to $ {\cal O}_{\m \n \r \s}$ some conditions must
be satisfied, i.e., it must have four Lorentz indices, its total canonical
dimension must be $n$ in units of mass, and must be finite when multiplying
it by $p^0$ and taking the limit of $p^0$ approaching infinity.
We will construct terms made up with combinations of ${\bf 1}$, $\th_{\m \n}$
and $\Th$ operators, and study their behavior. Other higher order
terms (e.g. $\Th^p$, with $p\in {\bf N}$) could be constructed, but they
will vanish when taking the BJL limit. The following are the possible terms:

\bs
\no i) The only operator appearing in the term is ${\bf 1}$.
In $n\ge2$ dimensions, the general expression for a term of such type
will be
\[    {1\over p^{4-n} }\; p_\m p_\n p_\r p_\s\;{\bf 1}. 	\label{7a}   \]
Except for the case $n=2$ which we studied in section III, this kind
of term fails to give a finite contribution when $p^0$ approaches infinity.

\bs
\no ii) The only operator is $\th_{\m\n}$. The expression for any value of $n$
will be of the form
\[      {p_\m p_\n \over {p^2}} \th_{\r\s},       \label{7b}    \]
which will only contribute to the commutator with a term like
\be
\eta_{\m (0} \eta_{\n i)} \th_{\r\s} \pa^i\d(\vec x). \label{7c}
\ee

\bs
\no iii) The only operator is $\Th$. For any value of $n$, the term will be
of the form
\[      {1\over p^4 }\; p_\m p_\n p_\r p_\s \Th           \label{7d}.     \]
In this case the commutator is proportional to
\[     \eta_{\m (0}  \eta_{\n 0} \eta_{\r 0} \eta_{\s i)} \;\Th\;
\pa^i\d(\vec x).  \label{7e}   \]

\bs
\no iv) Term made of products of the three operators: ${\bf 1}$,
$\th_{\m \n}$ and $\Th$. The only possibility is

\[   {1 \over {p^{n+2}}} (p_\m p_\n \th_{\r \s} + p_\r p_\s \th_{\m\n})\Th,  \]
which goes to zero as $p^0 \to \infty$. A term with more products of energy-
momentum tensors will require the existance of more momenta to compensate
for the extra indices, some of which will be the same as one of the
indices of the tensor, like in
\[   {1 \over {p^{2n+4}}} p_\m p_\n p_\a p_\b \th_{\r \s} \th^{\a\b} \Th.   \]
This term will be zero because we are considering only conserved tensors.
In any case, it will not contribute to the BJL limit, either.

We therefore see that there is no anomalous contribution to the
commutators for $n \ge 4$, as we wanted to show.

\begin{center}
\subsection*{V. Light-cone coordinates}
\end{center}

In the study of two dimensional systems it is usual to
work with light-cone (LC) coordinates. The LC picture is very physical and
also provides a useful framework for calculations. Historically, it was
LC quantization that first conclusevely established that dual models
were theories of strings \cite{super}.

We define the non-singular transformation
\ba
x^- &=& {1\over{\sqrt 2}} (x^0 - x^1)   \nn \\
x^+ &=& {1\over{\sqrt 2}} (x^0 + x^1). \label{lc}
\ea

In \cite{witten2} a ``canonical'' formalism was
proposed in which $x^-$
was interpreted as the ``time'' variable $x^0$, and $x^+$ as the ``space''
variable $x^1$. As we will show in this section the identification has to be
done carefully. The results in (\ref{7}) are Lorentz covariant and
valid for any inertial frame of reference. We encountered the anomaly coming
as a third derivative of the delta function, in $[\th_{01},\th_{11}]$. It
could appear, if one is not careful, that with the ``canonical'' formalism
proposed above, the anomaly should show up in the commutator
$[\th_{-+},\th_{++}]$, when we go to the light-cone frame.
In order to derive the well known anomalous
commutator given in references \cite{witten,guadagnini,super}, i.e.
$[\th_{++},\th_{++}]$, we will combine an OPE similar to the one in (\ref{5})
with the coefficients $c(p)$ written in terms of the light-cone
coordinates, and a generalization of the BJL limit for the LC coordinates.

Under (\ref{lc}) the components of the metric become
$\eta^{++}=\eta^{--}=0$ and $\eta^{+-}=\eta^{-+}= 1$.
If we denote by $x^c = (x^-,x^+)$ the position vector in the LC system,
and $P$ the matrix to transform from $x$ to $x^c$, i.e., $x^c = P x$, then the
energy-momentum tensor transforms according to
$T^c = P T P^{-1}$. This implies the following relations for the components
\ba
\th_{--}&=& \hf (\th_{00} - 2\th_{01} + \th_{11})	\nn	\\
\th_{-+} &=& \th_{+-} = \hf (\th_{00} - \th_{11})     \nn    \\
\th_{++} &=& \hf (\th_{00} + 2 \th_{01} + \th_{11}).    \label{9}
\ea

We can raise and lower indices with the help of the LC metric in the following
way $x_\pm = x^\pm$. Other results can be obtained by simple
algebraic manipulation
\ba
d^2 x &=& dx^0 dx^1 = \sqrt{2} dx^- dx^+    \nn  \\
px &=& p_- x^- + p_+ x^+     \nn   \\
p^2 &=& 2 p_- p_+,     \label{9a}
\ea
or in case of working with fermionic models, one could find the relations
\ba
&&\g^\pm = {1\over{\sqrt 2}}(\g^0 \pm \g^1) = \g_\pm  \nn  \\
&&(\g^+)^2 = (\g^-)^2 = 0   \nn  \\
&&\lbrace \g^+,\g^- \rbrace = 2.   \label{9b}
\ea

To evaluate the anomalous part of the commutators in the LC frame, we
will need to modify the expression for the BJL limit.
We can define the commutator in terms of LC coordinates as
\ba
\lim_{p_- \to \infty}\!\!\!\!&p_-&\!\!\!\!\int dx^+ dx^-
e^{i(p_+ x^+ + p_- x^-)} \langle\Om | T \,(\,\th_{\a\b} (x^-, x^+)
\th_{\e\d}(0)\,)\, |\Om\rangle =    \nn  \\
&&i \int dx^+ e^{i p_+ x^+} \langle\Om | \,[\,\th_{\a\b} (0, x^+),
\th_{\e\d}(0)\,]\,|\Om\rangle.   \label{LC}
\ea
Following the same steps as we did in the previous section, we now write the
OPE in terms of LC coordinates\footnote{Now the greek
indices $\a, \b, \e, \d$ take values $-, +$.}. We will only consider
those terms that will contribute when taking the
$p_-\to\infty$ limit, at the end. Let us consider
\be
{1\over{p_- p_+}}[ b_1 p_\a p_\b p_\e p_\d {\bf 1} + b_2 (p_\a p_\b
\th_{\e\d} + p_\e p_\d \th_{\a\b})],   \label{10}
\ee
multiply it by $p_-$ and finally take the limit of $p_-$ approaching
infinity. Note that when we take this limit in the covariant framework,
the denominator behaves as $(p^0)^2$ where as in
the LC frame it goes like $p_- p_+$. This causes the appearence of
off-diagonal terms of the metric $\eta_{-+}$ in the surviving coefficients,
once the LC-BJL limit is taken.
The final result is
\ba
[ \th_{\a\b}(0,x^+), \th_{\e\d}(0) ] &=& b_1 \eta_{-\a} \eta_{-\b}
\eta_{-\e} \eta_{-\d} \pa^3_+ \d(x^+)   \nn   \\
&-& b_2 (\eta_{-\a} \eta_{-\b} \th_{\e\d}  + \eta_{-\e} \eta_{-\d} \th_{\a\b})
\pa_+ \d(x^+).   \label{11}
\ea
The anomalous part is the term proportional to the third derivative
of the delta function. Since $\eta_{+-}=1$, the only non-vanishing
commutator appears for $\a=\b=\e=\d=+$. Therefore
\be
{[\th_{++}(0,x^+), \th_{++}(0)]}_{\rm Anom} = b_1 \pa^3_+ \d(x^+).
\label{virasoro}
\ee
This is in perfect agreement with the results found in the
literature\cite{witten,super} when working with LC coordinates.
If we were working with a non-traceless
energy-momentum tensor, there will be an extra contribution of
the form $ - b_3 \eta_{\a(+} \eta_{\b-} \eta_{\e-} \eta_{\d-)}\;\Th\;
\pa_+ \d(x^+)$, where again, the parentheses mean symmetric summation with
the second indices being switched around; in this case there appear four terms.

A particular scenario where one can find the result of eq. (\ref{virasoro})
is that concerning with Weyl spinors. The expression of the fermionic
energy-momentum tensor considered in \cite{witten} is the one given in
eq.(\ref{4a}). Since these spinors satisfy the Weyl relation
$ \g_{5} \psi = - \psi $, it is straightforward to verify other relations
like
\[	\g_{-} \psi = \pa_{-} \psi = 0.	\]
 From this, it can be shown that
\be
\th_{-+} = \th_{+-} = \th_{--} = 0,	\label{zero}
\ee
with $\th_{++}$ being the only non-vanishing component.

Comparing (\ref{7}) with the result obtained in
(\ref{virasoro}) using LC coordinates, there appear to be some kind of
contradiction if one is not careful.
 From (\ref{9}) and (\ref{zero}), ones finds
\[	\th_{00} = \th_{11} = \th_{01} = \th_{10} = {1\over 2}\th_{++}.	\]
which implies that $[\th_{00}, \th_{00}] = {1\over 4}
[\th_{++}, \th_{++}] \sim \d'''(x^+)$ from
(\ref{virasoro}), but $[\th_{00},\th_{00}] = 0$, from (\ref{7})!
The arguments of the delta functions in the Lorentz
covariant frame are $(x^0, x^1)$ and in the LC system are $(x^-, x^+)$,
so one needs to construct a simplectic form \cite{witten2} to go from one
to the other. The mere substitution using the expressions of (\ref{9}) when
going to the LC is not correct and a re-definition of the commutator
by means of the BJL-limit, as in eq.(\ref{LC}), needs to be
given in terms of the new coordinates. The results will then be consistent.
It has been previously noticed  that the LC formalism is
not manifestly covariant, althought it will be manifestly free of ghosts.
It is clear that the effect of Lorentz transformation on the coordinates
has to be quite subtle in the LC gauge since the choice of gauge is not
Lorentz invariant \cite{super}.

\begin{center}
\subsection*{VI. Jacobi Identity in two dimensions}
\end{center}

We have seen in sections III through V, by making an OPE and taking the
BJL limit, that the algebra of energy-momentum commutators presents
anomalies. It is also interesting to check if these anomalies that
appear in the single
commutators will violate the Jacobi identity for double commutators,
written as
\[	[[\th_{\m\n},\th_{\r\s}],\th_{\l\t}] +  [[\th_{\r\s},\th_{\l\t}],
\th_{\m\n}] + [[\th_{\l\t},\th_{\m\n}],\th_{\r\s}] =  0.   \]
The study of double commutators can be done in a similar manner to
the single ones. In section II we saw how the BJL limit allows us
to {\it define} commutators as the high energy behavior of correlation
functions. It should be quite natural to look for a double high energy limit
when interested in double commutators.

\bs
It is easy to verify the following relations
\ba
\frac{\pa}{\pa x_0} T A(x) B(y) C(z) &=& T(\dot A B C) +
\d(x_0 - y_0) T([A,B]_{(x_0)} C(z_0))  \nn   \\
&+& \d(x_0 - z_0) T(B(y_0)[A,C]_{(x_0)}),  \\
\frac{\pa}{\pa y_0} T A(x) B(y) C(z) &=& T(A B' C) +
\d(y_0 - x_0) T([B,A]_{(y_0)} C(z_0))  \nn   \\
&+& \d(y_0 - z_0) T(A(x_0)[B,C]_{(y_0)}),
\ea
where the dot and prime denote derivatives with respect to $x_0$ and
$y_0$ respectively. Following the same steps as in the derivation of
the traditional BJL limit (ref.\cite{Jackiwbook}), it can be
shown that if we define\footnote{Using the translational invariance
property of the Green's functions, we choose $z=0$ for simplicity.}
\[   {\cal O}(p,q) = \int d^n x d^n y e^{i (p x + q y)} \langle \a | T A(x)
B(y) C(0) |\b \rangle,  	\]
then the following identities hold
\ba
{\rm (I)}&=&\lim_{q_0 \to \infty} q_0 \lim_{p_0 \to \infty} p_0 \;
{\cal O}(p,q)\nn\\
&=&\int d^{n-1}x\; d^{n-1}y\; e^{-i(\vec p \cdot \x + \vec q \cdot \y)}
\langle \a |[ [ A(0,\x),C(0)],B(0,\y)]| \b \rangle \nn  \\
{\rm (II)}&=&\lim_{p_0 \to \infty} p_0  \lim_{q_0 \to \infty} q_0
\;{\cal O}(p,q) \nn    \\
&=&\int d^{n-1}x\; d^{n-1}y \;e^{-i(\vec p \cdot \x + \vec q \cdot \y)}
\langle \a |[ [ B(0,\y),C(0)], A(0,\x)]| \b \rangle    \nn    \\
{\rm (III)}&=&\lim_{k_0 \to \infty} k_0 \lim_{p_0 \to \infty} p_0
\;{\cal O}(p,q) \nn   \\
&=&\int d^{n-1}x\; d^{n-1}y\; e^{-i(\vec p \cdot \x + \vec q \cdot \y)}
\langle \a |[ [ A(0,\x), B(0,\y)], C(0)]| \b \rangle,  \label{2com}
\ea
where $k_0 = -(p_0 + q_0)$.

Thus the Jacobi identity will be given by ${\rm {- (I) + (II) + (III)}}$,
which is identically zero in absence of anomalies.

\bs
The behavior of gauge transformations in an anomalous gauge theory,
as well as in a consistent gauge theory with Chern-Simons term, can be given
a unified description in terms of cocycles. It is known that a 3-cocycle
arises when a representation of a transformation group is non associative,
and thus there is failure of the Jacobi identity \cite{3cocy1}. The
existance of these objects has been under investigation in quantum field
theory for some time now. In the context of quantum mechanics 3-cocycles
appear when translations are represented on configuration space $q$
in the presence of magnetic forces, specially a magnetic monopole. In
that case one finds that $J(v^1, v^2, v^3)= e \hbar^2/ m^3 \vec\nabla
\cdot \vec B$, where $v^i$ represent the components of the gauge invariant
velocity operator. If $\vec\na\cdot\vec B\not=0$, as in the case of a
point monopole\footnote{For a point monopole with strength $g$ located
at $\vec {r_0}$ the divergence of $\vec B$ is $\vec\na\cdot\vec B = 4\pi
g \d(\vec r - \vec {r_0})$.}, the Jacobi identity fails \cite{3cocy2}.

Also a violation of the Jacobi identity appears in the quark model. When
the Schwinger term in the commutator between time and space components
of a current is a c-number, the Jacobi identity for triple commutators
of spatial current components must fail \cite{3cocy3}. This fact has
been verified in perturbative BJL calculations \cite{3cocy4}. Also
non-perturbative calculations to find 3-cocycles associated to non-Abelian
gauge theories are under current investigation \cite{mine3}. There, the
use of OPE
techniques along with double BJL limits to study the Jacobi identity,
has proven to be an effective way to deal with the current algebra of
double commutators.

\bs
We would like to study the existance of 3-cocycles associated
to Abelian and Lie algebra valued energy momentum tensors. The study can
be easily restricted to the Abelian case once the non-Abelian one is known.
Thus the energy-momentum tensors will take values on a Lie algebra,
$\th_{\m\n}=\th^a_{\m\n} T_a$,
where the anti-hermitian representation matrices $T^a$ satisfy
the usual Lie algebra relation $[T^a, T^b]=f^{abc} T_c$, $(T^a)^\dagger=
-T^a$, and are normalized by ${\rm Tr}(T^a T^b)= -\d^{ab}$.

\bs
The definitions of eq.(\ref{2com}) are valid for any dimension, although
we will study here just the case of $n=2$. From the symmetries of the
energy-momentum tensor and those of the new Green's function itself
\be     {\cal O}^{abc}_{\m\n\r\s\l\t}(k_1,k_2) = \int d^2 x d^2 y
e^{i (k_1 x + k_2 y)} \langle \Om | T^* (\th^a_{\m\n}(x) \th^b_{\r\s}(y)
\th^c_{\l\t}(0))|\Om \rangle,  \label{double}
\ee
we find

\bs
\no i) Symmetry under the exchange $\m\lra\n $, $\r\lra\s$ or
$\l\lra\t$ separately.

\bs
\no ii) Symmetry under the exchange of pairs $(\m,\n) \lra (\r,\s)$,
$a\lra b$ and $k_1\lra k_2$.

\bs
\no iii) Symmetry under the exchange of $(\m, \n) \lra (\l, \t)$,
$a\lra c$, $k_1 \lra -(k_1 + k_2)$ and $k_2\lra k_2$.

\bs
\no iv) Symmetry under the exchange of $(\r, \s) \lra (\l, \t)$, $b\lra c$,
$k_1\lra k_1$ and $k_2 \lra -(k_1 + k_2)$.

\bs
 From (\ref{double}) we find the total canonical dimension of the opereator
to be $[{\cal O}]=({\rm mass})^2$.
To simplify the notation, we will not write all the symmetric partners
for each term in the OPE. However, symmetries i) --- iv) are understood
to be necessary. In
the operator expansion of (\ref{double}), we assume parity to be a
symmetry of the correlation function. Two possible candidate terms for
the OPE, which contribute to the double limits are
\be
u^{abc} k_{1\m} k_{1\n} k_{2\r} k_{2\s} k_{1\l} k_{2\t}{1\over{(k_1 + k_2)^2}}
({1\over{k_1^2}}+{1\over{k_2^2}})\;{\bf 1}.  \label{jacobi1}
\ee
and
\be
{w^{abc}}_d \; k_{1\m} k_{1\n} k_{2\r} k_{2\s} {1\over{(k_1+k_2)^2}}
({1\over{k_1^2}} + {1\over{k_2^2}})\;\th^d_{\l\t},   \label{jacobi2}
\ee
with their properly symmetrized partners. In (\ref{jacobi1}) the function
$u^{abc}$ is made of combinations of the Lie algebra totally antisymmetric
and symmetric tensors $f^{abc}$ and $d^{abc}$. Therefore, we can write
$u^{abc} = u_1 f^{abc} + u_2 d^{abc}$. Similar arguments hold for
$w$ which will be of the form ${\rm Tr}(T^a T^b T^c T^d$.
Detailed calculations can be carried out
to evaluate the three double commutators and for both terms we find the
Jacobi identity not to be violated.

If the energy-momentum tensor we are dealing with is not traceless, then
there is another possible term to appear in the OPE. This will be of
the form
\be
z^{abc} k_{1\m} k_{1\n} k_{2\r} k_{2\s} k_{1\l} k_{2\t} ({1\over{k_1^4
k_2^2}} + {1\over{k_1^2 k_2^4}})\;\Th,
\ee
and its symmetric partners, which satisfy relations i) --- iv). The
explicit evaluation of the double commutators in expression (\ref{2com})
leads again to the no violation of the Jacobi identity.

In two dimensions, as we mentioned in section IV, is
quite convenient the use of LC coordinates. A straightforward generalization
of the method shown above for the expression of double commutators, can
also be obtained in terms of LC coordinates.

\begin{center}
\subsection*{VII. Conclusions}
\end{center}

In this paper we considered the BJL definition for commutators
and applied it in conjunction with the operator product expansion. The
method could be applied both in the perturbative and non-perturbative
regimes. We first derived the covariant expression for commutators of
energy-momentum tensors in $n=2$ dimensions, and found anomalous terms
coming from third derivatives of delta functions. This anomalous result
appeared again when using the BJL technique in the light-cone frame, and
was in perfect agreement with those found in the literature. The method
allowed us to calculate possible anomalous terms in other dimensions in a very
straightforward way. In section IV, in particular, we found that there
are no anomalous commutators in even dimensionsal Minkowskian spacetime for
$n\not=2$.

The method requires some additional knowledge about the behavior of the
coefficient funtions (which appear in the OPE) at large momentum transfer
$p$. In asymptotically free theories this is available via the renormalization
group. The final results are expressed in terms of the residues of the
coefficient functions (i.e.; the constant multiplying the term behaving as
$1/{p^0}$) and of the matrix elements of various local operators (the
``condensates''). These quantities can be evaluated explicitly within
perturbation theory; in the non-perturbative regime the condensates
cannot be evaluated explicitly but can be used to parametrize the
results. Another characteristic of the method is that the results
are evaluated in terms of a set of unknown constants, the residues of the
coefficient functions $c_i$. For the applications we have considered
this will not be a disadvantage, since we are just interested in the anomalous
behavior of the energy-momentum commutators. The exact coefficients for
these terms can only be found through detailed explicit calculation
for each matrix element.

The existence of these anomalies at the single commutator level lead us
to study the possibility of violations of the Jacobi identity. A
generalization of the BJL limit technique to a double limit \cite{mine3},
allows us
to find expressions for double commutators, and therefore be able to
check whether or not the Jacobi identity holds. It vanishes at the
canonical level, but when
higher loop corrections are brought in, there is hope of generating
3-cocycles, as in other field theoretical contexts. From the terms
studied in section VI, we conclude that there are no
violations in the Jacobi identity of energy-momentum tensors in two
dimensions, at the Abelian and non-Abelian level.
This result concerns only with the non-interacting models
(fermionic or bosonic) in a flat background metric. We have not studied
the Jacobi identity for the case of $n\not=2$, or the possibility of
having mixed anomalies. These examples can be studied in the same way as
we have done for $n=2$, and the generalization is straightforward.

\bs
\subsection*{Acknowledgements}

J.P.M acknowledges helpful discussions with Jos\'e Wudka.
This work was supported in part through funds provided by the Department
of Energy under contract DE-FG03-94ER40837.

\vfill

\ed
\begin{thebibliography}{10}

\bibitem{Schw} J.\ Schwinger, {\it Phys.\ Rev.\ Lett.\ } {\bf 3}, 269 (1959).
T.\ Goto and I.\ Imamura, {\it Prog.\ Theor.\ Phys.\ } {\bf 14}, 196 (1955).

\bibitem{BJL} J.\ D.\ Bjorken {\it Phys.\ Rev.\ } {\bf 148}, 1467 (1966).
K.\ Johnson and F.\ E.\ Low {\it Prog.\ Theor.\ Phys.\ } {\bf Suppl.\ 37-38},
74 (1966).

\bibitem{Wilson} K.\ G.\ Wilson, {\it Phys.\ Rev.\ }{\bf 179} 1499 (1969).

\bibitem{Jo} To see an example of how involved perturbative calculations can
be, S.\ Jo {\it Nuc.\ Phys.\ } {\bf B259}, 616 (1985).

\bibitem{mine1} J.\ P.\ Muniain and J.\ Wudka, Report UCRHEP-T132.

\bibitem{Jackiwbook} For a review on the BJL limit see
S.\ Treiman, R.\ Jackiw, B.\ Zumino, E.\ Witten,
{\it Current Algebra and Anomalies}, Princeton University Press, Princeton,
N. J., 1985.

\bibitem{witten} L.\ Alvarez-Gaum\'e and E.\ Witten, {\it Nuc.\ Phys.\ }
{\bf B234}, 269 (1983).

\bibitem{guadagnini} E.\ Guadagnini, {\it Phys.\ Rev.\ }{\bf D38} 2482 (1988).

\bibitem{super} M.\ B.\ Green, J.\ H.\ Schwarz, E.\ Witten, {\it Superstring
theory}, Cambridge University Press, Cambridge, 1987.  \\
Include also the original reference for Virasoro's paper

\bibitem{mine2} A.\ N.\ Redlich, {\it Phys.\ Rev.\ Lett.\ }{\bf 52}, 18 (1984);
{\it Phys.\ Rev.\ }{\bf D29}, 2366 (1984). Javier P.\ Muniain, {\it Anomalous
current commutators in $1+2$-dimensions}, work in progress.

\bibitem{witten2} E.\ Witten, {\it Commun.\ Math.\ Phys.\ }
{\bf 92}, 455 (1984).


\bibitem{3cocy1} R.\ Jackiw, {\it Phys.\ Rev.\ Lett.\ }{\bf 54}, 159 (1985);
{\it Phys.\ Rev.\ Lett.\ }{\bf 54}, 2380 (1985). J.\ Mickelsson, {\it Phys.\
Rev.\ Lett.\ }{\bf 54}, 2379 (1985).

\bibitem{3cocy2} For a brief review on cochains, cocycles and the coboundary
operation in quantum mechanics and field theory, see reference
\cite{Jackiwbook}.

\bibitem{3cocy3} F.\ Bucella et al, {\it Phys.\ Rev.\ }{\bf 149}, 1268 (1965).

\bibitem{3cocy4} K.\ Johnson and F.\ E.\ Low {\it Prog.\ Theor.\ Phys.\ }
{\bf Suppl.\ 37-38}, 74 (1966).

\bibitem{mine3} J.\ P.\ Muniain and J.\ Wudka, {\it 3-Cocycles and the operator
product expansion}, work in progress.

\end{thebibliography}
